\documentclass[utf8]{FrontiersinHarvard}

\usepackage{url,hyperref,microtype}
\usepackage[onehalfspacing]{setspace}
\usepackage{booktabs}
\usepackage{multirow}
\usepackage{tabularx}
\usepackage{textcomp}
\usepackage{xcolor,soul}
\newif\ifshowrevisions
\showrevisionsfalse
\sethlcolor{yellow!25}

\DeclareRobustCommand{\revmath}[1]{\ifshowrevisions\colorbox{yellow!25}{$#1$}\else$#1$\fi}

\graphicspath{{figures/}}
\newsavebox{\modeloverview}
\hypersetup{hidelinks=true}

\def\keyFont{\fontsize{8}{11}\helveticabold}
\def\firstAuthorLast{Zhang {et~al.}}
\def\Authors{Yaowen Zhang\,$^{1}$, Shuhao Que\,$^{1}$ and Ying Wang\,$^{1,*}$}
\def\Address{$^{1}$Department of Biomedical Signals and Systems, University of Twente, Enschede, The Netherlands}
\def\corrAuthor{Ying Wang}
\def\corrEmail{imwywk@gmail.com}

\begin{document}
\onecolumn
\firstpage{1}

\title[]{PMB-NN: Physiological-Model-Based Neural Network for Glucose Forecasting in Type 1 Diabetes}
\author[\firstAuthorLast]{\Authors}
\address{}
\correspondance{}
\extraAuth{}

\maketitle

\begin{abstract}
\section{}
Accurate, physiologically consistent glucose forecasts can support earlier responses to glucose changes in type 1 diabetes. We adapted our physiological-model-based neural network (PMB--NN) framework by jointly training a temporal convolutional network using 180 min of history with a reduced physiological glucose model containing four personalized parameters. The main two-term formulation guided learning through neural forecast error and disagreement between neural and physiological forecasts. To examine the benefit of directly fitting physiological predictions to measured glucose, we tested a three-term extension with this additional loss and a reduced consistency weight. Evaluation used 6--8 weeks of continuous glucose monitoring, insulin-pump, and carbohydrate records from 25 AZT1D participants. Participant-specific chronological 70/10/20 training, validation, and test splits produced six settings: predictions 30, 60, and 120 min ahead, each with or without available basal-insulin records. Numerical accuracy was compared with neural benchmarks and the standalone physiological model. Plausibility tests examined whether predicted glucose was lower or unchanged after added insulin and higher or unchanged after added carbohydrate. Personalized parameters described the glucose operating point, return rate, and insulin- and carbohydrate-response strengths; separate analyses assessed parameter distinguishability and estimation stability. Median mean absolute errors of the two-term model were 16.24, 25.00, and 30.20 mg/dL with basal records and 15.38, 24.21, and 31.79 mg/dL without them at the respective prediction times. It achieved the lowest median error at 30 and 60 min without basal records and outperformed the standalone physiological model in all six settings. Median expected-direction response rates were 85\% for additional insulin at high glucose and 100\% for additional carbohydrate at low glucose. Independent physiological fits gave more stable estimates of the operating point and return rate than of input-response strengths. Compared with the two-term model, the three-term extension did not significantly improve overall error and produced less consistent physiological responses, despite lower errors during basal-context high-glucose periods. Overall, PMB–NN supports personalized glucose forecasting while making physiological input responses and parameter evidence explicit, potentially contributing to interpretable early-warning tools for individualized diabetes management.

\tiny
\keyFont{\section{Keywords:} type 1 diabetes, glucose forecasting, physiological-model-based neural network, hybrid artificial intelligence, physiological plausibility, personalization}
\end{abstract}

\section{Introduction}
\label{sec:introduction}

Continuous glucose monitoring (CGM) and automated insulin delivery (AID) systems provide frequent glucose and insulin records, alongside logged carbohydrate intake, for people with type 1 diabetes (T1D). Glucose forecasting uses these records to anticipate rising or falling glucose before an adverse level is reached, providing potential lead time for a warning and a response within an individual's care plan \citep{danne2017international,marling2020ohiot1dm,martinsson2020blood}. Forecasts must account for delayed insulin action, variable meal absorption, and subsequent activity or treatment that is unknown at prediction time. A useful personalized model should therefore combine numerical accuracy with responses to insulin and carbohydrate that follow their expected physiological effects.

Data-driven sequence models learn nonlinear temporal patterns from CGM histories and can capture variation beyond a small set of physiological equations \citep{martinsson2020blood,bai2018empirical}. However, minimizing forecast error alone does not enforce the expected glucose responses to insulin and carbohydrate, and neural-network weights do not directly describe individual physiological properties. Learned insulin--glucose associations can also reflect treatment decisions, because an AID system may increase insulin delivery when glucose is already rising. Physiological models explicitly represent glucose regulation, insulin action, and meal effects through interpretable parameters \citep{bergman1979quantitative,hovorka2004nonlinear,dallaman2007meal}. Their limitations arise from simplified equations and the difficulty of estimating hidden states and parameters from routine measurements. Reduced models can miss changes associated with activity, stress, or variable absorption, limiting their ability to fit free-living glucose dynamics. These complementary strengths and limitations motivate combining the two approaches: physiological relationships can guide neural-network training and provide interpretable parameters, while the neural component retains the flexibility to capture patterns not represented by the reduced equations.

As one type of the state-of-art hybrid AI frameworks, our earlier proposed physiological-model-based neural network (PMB-NN) framework trains data-driven models constrained by relevant reduced physiological models for either heart-rate or blood-pressure estimation \citep{zhang2026PMBNN1,zhang2026PMBNN2}. Both of  the variant PMB-NN models achieve comparable accuracy with the pure data-driven models and meanwhile outperform the data-driven models in the aspect of physiological plausibility and interpretability. Accordingly, we further proposed a PMB-NN family’s variant model for the glucose forecasting using a temporal convolutional network (TCN) and a reduced glucose--insulin--carbohydrate model. The main two-term formulation combines a data-fitting loss with a loss encouraging agreement between the data-driven and physiological forecasts. A three-term extension additionally fits the physiological forecast to measured glucose, as examined in our related simultaneous-learning work \citep{Que2026HybridPhysicsAI}. We assess numerical accuracy, physiological plausibility, and parameter interpretability, and compare the two formulations in both their outcomes and optimization behavior. Short-duration calibration further examines the amount of personal data needed for forecasting for the consideration of real-world adoption in future.

 \section{Materials and methods}
\label{sec:method}

\subsection{Dataset and preprocessing}
\label{sec:dataset_preprocessing}

\subsubsection{Dataset}
\label{sec:dataset}

We used the AZT1D dataset of free-living AID records from 25 people with T1D (12 male, 13 female) \citep{khamesian2025azt1d}. Median age was 65 years (range 27--80), and median HbA1c was 6.7\% (range 5.0\%--8.2\%). Each recording covered approximately 6--8 weeks; together the records contained 26,707 h of CGM, insulin-pump, and carbohydrate data.

We separated each recording according to basal-data availability. The earlier no-basal context comprised periods without recorded basal delivery in 24 participants (median duration 16.3 days, range 6.8--20.8); the later basal context comprised periods with reliable basal records in all 25 participants (median 27.8 days, range 25.8--27.8). Each context was then divided chronologically into training, validation, and test segments.

\subsubsection{Preprocessing}
\label{sec:preprocessing}

CGM was placed on a 5-min grid. Total bolus insulin delivered was taken as a dose in units, and carbohydrate size was taken as grams. Pump mode was excluded from the model inputs. Basal availability was preserved separately in the two recording contexts. Each recorded hourly basal rate (U/h) was assigned to its valid 5-min intervals for at most 75 min, and the delivered dose per interval was calculated as $D_{\mathrm{basal}}(t)=\mathrm{BasalRate}(t)\,5/60$ U. Basal-context samples required valid basal information throughout their input history and forecast targets. In no-basal samples, a zero basal channel encoded unavailable delivery records; actual basal delivery during that period was unobserved.

We forecast glucose 30, 60, and 120 min ahead of each current CGM observation; these intervals are termed forecast horizons. A sample was retained only when current glucose and all three future measurements were available. The first two horizons reflect potential short-term warnings; the 120-min horizon probes how far predictions can extend without knowing subsequent meals or insulin delivery \citep{marling2020ohiot1dm,danne2017international}.

Each neural-network sample contained 36 consecutive 5-min steps (a 180-min input window). At each step, the nine inputs were
\begin{equation}
\begin{aligned}
\bigg[
& G,\, D_{\mathrm{bolus}},\, D_{\mathrm{basal}},\, C,\, \mathrm{IOB}_{\mathrm{bolus}},\, \mathrm{IOB}_{\mathrm{basal}},\, \mathrm{COB}, \\
& \sin(2\pi h/24),\, \cos(2\pi h/24)
\bigg],
\end{aligned}
\end{equation}
where $G$ is CGM, $D_{\mathrm{bolus}}$ and $D_{\mathrm{basal}}$ are insulin doses, $C$ is carbohydrate intake, IOB and COB are derived active-input states defined below, and $h$ is clock hour. The sine and cosine terms keep nearby times around midnight close in the input representation.

\subsection{PMB--NN model for glucose forecasting}
\label{sec:architecture}
\begin{figure*}[t]
\centering
\includegraphics[width=\textwidth]{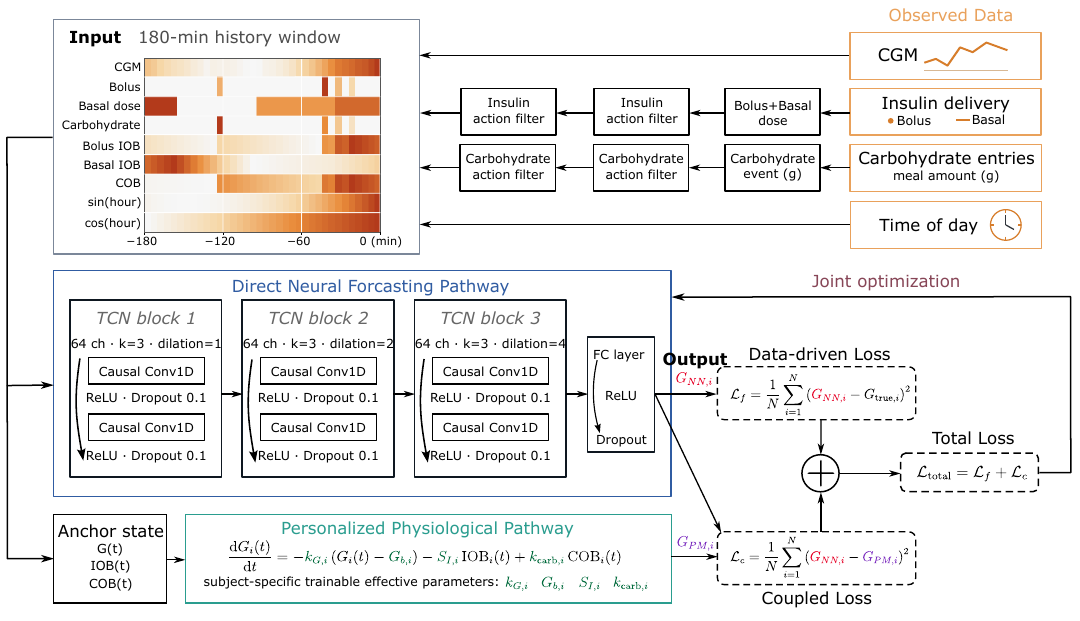}
\caption{Workflow of the main two-term glucose PMB--NN. Data preparation produces a 36-step history of CGM, insulin, carbohydrate, derived action states, and time of day. The neural module contains three causal TCN blocks and forecasts glucose at 30, 60, and 120 min. The physiological module uses current glucose, insulin on board (IOB), and carbohydrate on board (COB) to forecast the same prediction times. The two modules are trained jointly using neural forecast-to-reference error ($L_f$) and neural-to-PM disagreement ($L_c$); the TCN supplies the final output. The three-term extension adds PM forecast-to-reference error ($L_p$) with the weights specified in Eq.~\eqref{eq:three_loss_methods}. CGM, continuous glucose monitoring; TCN, temporal convolutional network; PM, physiological model; PMB--NN, physiological-model-based neural network; $G_b$, glucose operating point; $k_G$, return rate; $S_I$, insulin-response coefficient; $k_{\mathrm{carb}}$, carbohydrate-response coefficient.}
\label{fig:pmb_tcn_architecture}
\end{figure*}

The glucose PMB--NN consists of a data-driven TCN and a reduced physiological model (Fig.~\ref{fig:pmb_tcn_architecture}), extending our earlier framework \citep{zhang2026PMBNN1,zhang2026PMBNN2}. Both generate forecasts at the same horizons so that their predictions can be compared directly during training. The main two-term loss combines the error between the TCN forecast and measured glucose with the disagreement between TCN and physiological forecasts. The three-term extension adds the error between the physiological forecast and measured glucose. The TCN provides the final glucose prediction; the physiological model guides its training through the agreement loss.

\subsubsection{Reduced personalized physiological branch}
\label{sec:mathematical_model_methods}

We retained three physiological effects that can be represented using the available CGM, insulin, and carbohydrate records: return toward an operating glucose level, glucose lowering by active insulin, and glucose raising by available carbohydrate \citep{bergman1979quantitative,hovorka2004nonlinear,dallaman2007meal}. This reduction avoids estimating unmeasured plasma-insulin and meal-absorption states while allowing four participant-specific parameters to be fitted. We assumed a linear return term, additive insulin and carbohydrate contributions, fixed action filters, and parameters that remain constant within each participant--context recording. These assumptions define an effective input--output model for the observed period. For participant $i$, the model was
\begin{equation}
\frac{dG_i(t)}{dt}
=
-k_{G,i}\left(G_i(t)-G_{b,i}\right)
-S_{I,i}\mathrm{IOB}_i(t)
+k_{\mathrm{carb},i}\mathrm{COB}_i(t).
\label{eq:additive_phys_methods}
\end{equation}
Here $G_{b,i}$ is an effective glucose operating point, $k_{G,i}$ is the return rate, $S_{I,i}$ is the glucose-lowering coefficient of active insulin, and $k_{\mathrm{carb},i}$ is the glucose-raising coefficient of available carbohydrate. These four coefficients correspond to the operating level, intrinsic return dynamics, insulin effect, and carbohydrate effect retained in the reduced equation. Each participant's coefficients were estimated separately. For the standalone physiological-model (PM) forecasting comparator, the same equation was fitted directly to measured future glucose using AdamW with learning rate $5\times10^{-3}$, bounded parameters, and no weight decay. Validation RMSE selected its checkpoint under the same training schedule as the main experiment.

Insulin on board (IOB) and carbohydrate on board (COB) represent recent doses that remain active in the model. At each 5-min step, bolus and basal IOB were updated separately as $\mathrm{IOB}_{x}(t+5)=e^{-5/240}\mathrm{IOB}_{x}(t)+D_x(t)$ for $x\in\{\mathrm{bolus},\mathrm{basal}\}$; COB was updated as $\mathrm{COB}(t+5)=e^{-5/120}\mathrm{COB}(t)+C(t)$. The 240- and 120-min decay constants approximate insulin action and mixed-meal absorption \citep{home2012pharmacokinetics,dallaman2006system}. Total IOB was the sum of the two insulin states in basal context and bolus IOB alone in no-basal context. These derived states summarize delivery history in units and grams using fixed action filters.

The four parameters were kept in prespecified ranges with a bounded sigmoid transformation:
\begin{align}
70 \leq G_{b,i} \leq 250,\qquad
0.001 \leq k_{G,i} \leq 0.03,\\
0 \leq S_{I,i} \leq 2.0,\qquad
0 \leq k_{\mathrm{carb},i} \leq 0.08.
\end{align}
Units are mg/dL for $G_b$, min$^{-1}$ for $k_G$, mg/dL/min/U for $S_I$, and mg/dL/min/g for $k_{\mathrm{carb}}$. The bounds allow variation in observed glucose levels and response strengths while preserving the expected signs of insulin and carbohydrate effects. The fitted coefficients describe effective responses within this reduced model and its fixed action filters.

The initial $G_b$ was the median training CGM clipped to its allowed range. Initial values for $k_G$, $S_I$, and $k_{\mathrm{carb}}$ were 0.006, 0.12, and 0.015 in their respective units. Each participant's parameters were fitted from that participant's training windows.

Starting at the current CGM value, the physiological model projected glucose forward using IOB and COB at the prediction time and their fixed decay. With $b=k_G$ and $a=k_GG_b-S_I\mathrm{IOB}+k_{\mathrm{carb}}\mathrm{COB}$ held constant within each 5-min interval, the exact step was
\begin{equation}
G(t+\Delta t)=\frac{a}{b}+\left[G(t)-\frac{a}{b}\right]e^{-b\Delta t},
\label{eq:exact}
\end{equation}
Repeating this update generated physiological forecasts at 30, 60, and 120 min. We denote the resulting three-value forecast by $\widehat{\mathbf G}_{\mathrm{PM}}$. Before computing losses, it was normalized with the same training-set glucose statistics as the observed targets.

\subsubsection{Temporal convolutional network}
\label{sec:pmb_tcn_methods}
The TCN received the $36\times9$ input window defined above. Three residual causal blocks used 64 channels, kernel size 3, dilations 1, 2, and 4, and dropout 0.1 \citep{bai2018empirical}. Each block had two convolutions and a residual connection. This dilation sequence gave the final time step a 29-step receptive field, covering the most recent 145 min of the supplied 180-min history. A 64-unit dense layer and three-output head produced the normalized neural forecast $\widehat{\mathbf G}_{\mathrm{NN}}$ for the three horizons.

The reported PMB--NN forecast was the direct neural output:
\begin{equation}
\widehat{\mathbf{G}}_{\mathrm{PMB}}(t)
=
\widehat{\mathbf{G}}_{\mathrm{NN}}(t)
=f_{\mathrm{TCN}}\left(\mathbf{X}_{t-35:t}\right).
\label{eq:pmb_tcn_output_methods}
\end{equation}

\subsubsection{Physiological-model constraints during training}
Following our PMB--NN framework \citep{zhang2026PMBNN1,zhang2026PMBNN2}, the main objective combined data fitting with a physiological-model constraint:
\begin{align}
L_f&=\mathrm{MSE}\!\left(\widehat{\mathbf G}_{\mathrm{NN}},\mathbf G\right),\\
L_c&=\mathrm{MSE}\!\left(\widehat{\mathbf G}_{\mathrm{NN}},\widehat{\mathbf G}_{\mathrm{PM}}\right),
\end{align}
The data loss $L_f$ compares the neural prediction with measured future glucose. The physiological-model constraint $L_c$ compares the neural prediction with the PM prediction. Both receive unit weight in the main two-term formulation:
\begin{equation}
\mathcal L_{2\mathrm L}=L_f+L_c.
\label{eq:pmb_loss_methods}
\end{equation}
The same $L_c$ updates both the TCN and the personalized PM parameters. Thus the two modules are learned together, while only the TCN prediction is reported.

Both modules were optimized jointly with AdamW, batch size 512, gradient clipping at 5, and up to 60 epochs. TCN weights used learning rate $10^{-3}$ and weight decay $10^{-4}$; PM parameters used learning rate $5\times10^{-3}$ without weight decay. The checkpoint with the lowest mean validation RMSE across the three horizons was retained. Test data did not enter checkpoint or hyperparameter selection.

For the three-term PMB--NN extension, we also fitted the physiological prediction to measured glucose:
\begin{equation}
L_p=\mathrm{MSE}\!\left(\widehat{\mathbf G}_{\mathrm{PM}},\mathbf G\right),
\end{equation}
The extension objective was
\begin{equation}
\mathcal L_{3\mathrm L}=L_f+0.5L_c+L_p.
\label{eq:three_loss_methods}
\end{equation}
The architecture, neural output, initialization, and optimizer were otherwise the same. The comparison evaluates the joint change of adding direct PM supervision through $L_p$ and reducing the consistency weight from 1 to 0.5.

\subsection{Experimental setting}
\label{sec:experiments}

\subsubsection{Personalized chronological split}
\label{sec:train_test_split}

For each participant and recording context, windows were ordered by time. The first 70\% were used for training, the next 10\% for validation, and the final 20\% for testing. Windows were removed at split boundaries if their 120-min target crossed into the next partition. All models used identical windows and glucose targets. The validation set selected model checkpoints; all reported outcomes used the test set, i.e., the final 20\% of windows excluded from training and checkpoint selection.

\subsubsection{Short-duration calibration setting}

We also tested the main two-term PMB--NN with 1, 3, 6, 10, or 15 days of personal data in basal context. All 25 basal-context records supported these five durations. Only 3 of 24 no-basal records had 15 pre-test days, so no-basal records were excluded from this paired comparison. The full pre-test period had a median duration of 22.2 days (range 20.5--22.5).

Each duration used the first contiguous $D$ days of the original pre-test period, with the same test windows as the full-data model. Short-duration fits retained the final epoch of a fixed 60-epoch schedule, using all selected windows for training. The full-data protocol used the original 70\% training and 10\% validation partitions with validation-selected checkpoints. The comparison therefore evaluates the complete calibration protocols, including both data duration and checkpoint selection.

\subsubsection{Comparator models}

Data-driven comparators received the same $36\times9$ histories, targets, and normalization. They were an identical TCN trained without the PM constraint, a gated recurrent unit (GRU), a long short-term memory network (LSTM), and a transformer. The recurrent models had two 64-unit layers; the transformer had three encoder layers, four attention heads, 64 hidden units, and 256 feed-forward units. The standalone PM used the same anchors and targets, with current glucose and derived action states as its inputs.

\subsection{Model evaluation}
\label{sec:model_evaluation}
Following our earlier PMB--NN studies \citep{zhang2026PMBNN1,zhang2026PMBNN2}, we assessed numerical accuracy, physiological plausibility, and the physiological interpretation of the fitted parameters. Parameter interpretation was based on the processes represented in the reduced equation; separate sensitivity and resampling analyses examined local identifiability and estimation stability. We then used these outcomes to compare the two training objectives and examined how much personal data were needed for calibration. 

\subsubsection{Numerical accuracy}

Error metrics were calculated separately for each participant, context, horizon, and model. Mean absolute error (MAE, mg/dL) was the primary measure; RMSE, $R^2$, and upper-tail absolute errors provided additional measures of accuracy. Cohort medians summarized participant-level results without assigning greater weight to longer records. For the loss-formulation comparison, we also calculated MAE for targets with observed glucose above 180 mg/dL, defined here as hyperglycaemia.

Paired two-sided Wilcoxon signed-rank tests compared PMB--NN with each comparator within a context and horizon, followed by Holm correction across comparators \citep{wilcoxon1945individual,holm1979simple}. A nonsignificant result was not treated as proof of equal performance. For error differences, positive comparator-minus-PMB--NN values favor PMB--NN.

\subsubsection{Physiological plausibility}

Physiological plausibility was evaluated through four controlled input changes at prediction times in the test set. Increased active insulin and a correction bolus at glucose $\geq180$ mg/dL were expected to decrease or leave unchanged the forecast. Increased active carbohydrate and a rescue carbohydrate entry at glucose $\leq80$ mg/dL were expected to increase or leave it unchanged. We changed the relevant input and its derived action state and compared the resulting forecast with the original. Tested increases were 0.5--2 U for active insulin or bolus, 10--40 g for active carbohydrate, and 5--20 g for rescue. For each participant, context, horizon, and input change, the direction-plausible fraction was the number of test windows with the expected response divided by the number of eligible windows. A value of 1 means that every tested window followed the expected direction; 0.5 means that half did. Response magnitude was also calculated because an unchanged forecast can satisfy the direction criterion. These perturbations characterize model behavior.

For Fig.~\ref{fig:counterfactual}e, we selected the largest tested dose for each input change and took the median of the participant-level fractions across recording contexts and forecast horizons. Each displayed model therefore has four summary fractions, one for each input-change type. In Fig.~\ref{fig:counterfactual}f, the vertical coordinate is the arithmetic mean of those four fractions, and the horizontal coordinate is the median participant-level 30-min MAE across the two contexts. Lower error and a larger fraction of expected responses place a model toward the upper left. The dashed line links the displayed models for which no other comparator combines both lower MAE and higher mean direction plausibility. This plot summarizes the accuracy--response tradeoff; it is not an additional significance test.

\subsubsection{Physiological parameter interpretation and reliability}

The physiological interpretation comes from the roles assigned to the four parameters in Eq.~\eqref{eq:additive_phys_methods}: $G_b$ specifies the glucose operating point, $k_G$ the return rate, $S_I$ the glucose-lowering strength of IOB, and $k_{\mathrm{carb}}$ the glucose-raising strength of COB. We reported each person's PMB--NN estimates and their cohort distributions. Separate diagnostic analyses then assessed whether the available records supported locally distinguishable parameter effects and stable estimates, providing evidence about the reliability of this model-based interpretation.

To obtain an independent reference for the jointly learned coefficients, we also fitted the same physiological equation directly to measured training glucose using bounded nonlinear least squares with SciPy's trust-region reflective solver and a soft-L1 objective. This diagnostic fit was separate from the AdamW-trained PM forecasting comparator. It allowed us to compare parameters learned through agreement with the TCN in two-term PMB--NN with parameters fitted directly to observations. Agreement between the two estimates describes how similarly the same reduced equation was fitted under these different targets.

At each independent diagnostic solution, a sensitivity Jacobian measured forecast changes after small perturbations of the four physical parameters. It stacked the 30-, 60-, and 120-min forecasts over selected training windows, using parameters scaled by their allowed ranges and a fixed scale calculated from observed training glucose targets. Full numerical rank indicates distinguishable first-order local parameter effects; a larger condition number indicates more unequal sensitivities along combinations of the scaled parameters. Rank robustness was checked at three perturbation step sizes. We then refitted the independent PM to 100 resamples of contiguous 24-h training blocks to assess variability across sampled recording days \citep{efron1994introduction}. We reported 95\% percentile intervals and the proportion with coefficient of variation (CV) below 0.5. PMB--NN point estimates were compared with these independent-PM intervals to assess estimator agreement. Finally, Spearman correlations measured consistency of PMB--NN participant rankings between contexts. Full calculation details are given in Supplementary Methods.

\subsubsection{Comparison of the two training objectives}
\label{sec:loss_formulation_eval}

The two- and three-term PMB--NN models used the same windows, architecture, initialization, validation rule, and reported neural output. We compared their accuracy, glucose-range errors, controlled input responses, and fitted parameters using paired subject-level tests. Holm correction was applied within each family of comparisons.

To understand how the losses interacted during training, we measured the cosine similarity of their minibatch gradients \citep{yu2020gradient}. A value below zero means that the two losses would move the same weights in opposing directions. For the neural weights, both formulations have gradients from $L_f$ and $L_c$, so we compared those two gradients. For the physiological parameters in the three-term formulation, we compared $L_p$ with $L_c$. The other cross-pairs do not share a parameter group: $L_f$ has no physiological-parameter gradient, and $L_p$ has no neural-weight gradient.

\begin{equation}
\gamma_{fc}=\cos(\nabla_{\boldsymbol\theta}L_f,\nabla_{\boldsymbol\theta}L_c),
\qquad
\gamma_{pc}=\cos(\nabla_{\boldsymbol\phi}L_p,\nabla_{\boldsymbol\phi}L_c),
\end{equation}
where $\boldsymbol\theta$ denotes neural weights and $\boldsymbol\phi$ the physiological parameters. The second cosine is defined only for the three-term model. We summarized minibatch cosines by epoch; full statistical families and additional outcomes are described in Supplementary Methods.

\subsubsection{Short-duration calibration evaluation}

For each basal-context participant, the five short-duration models and the full-data model were evaluated on identical test times. Paired two-sided Wilcoxon tests compared MAE across the 15 duration--horizon combinations, with Holm correction over all 15 tests. Participant medians and interquartile ranges summarized accuracy. We also calculated short-duration-to-full-data MAE ratios and compared parameter estimates using absolute relative differences and participant-rank correlations. The core accuracy results are presented in Table~\ref{tab:calibration}; participant variation and parameter comparisons are shown in Supplementary Fig.~S1.

\section{Results}
\label{sec:results}

\subsection{Numerical accuracy}

\begin{figure*}[t]
\centering
\includegraphics[width=\textwidth]{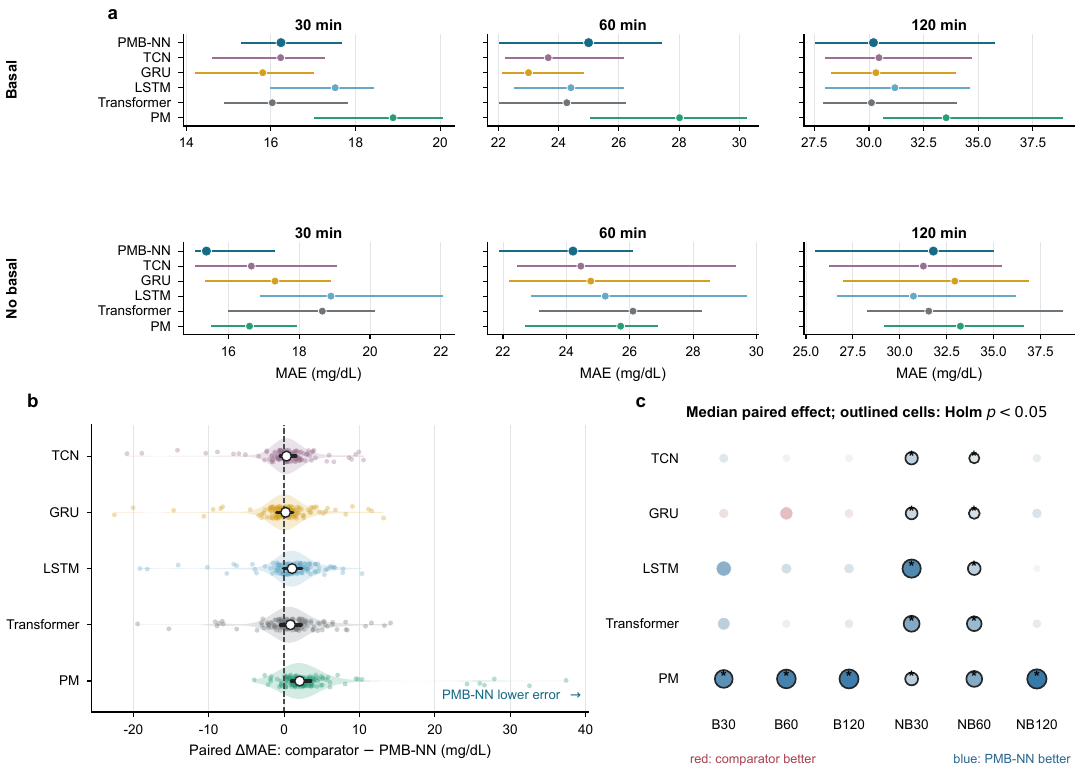}
\caption{Numerical accuracy of the two-term PMB--NN} and the comparator models. (a) Participant-median MAE with 95\% bootstrap confidence intervals for each recording context and forecast horizon. (b) Distributions of paired MAE differences pooled across participants and settings; black segments and open circles indicate the interquartile range and median. The horizontal difference is comparator MAE minus PMB--NN MAE: positive values to the right indicate lower error for PMB--NN, negative values to the left indicate lower error for the comparator, and zero indicates equal error. (c) Median paired differences within each setting. B30, B60, and B120 denote basal-context forecasts at 30, 60, and 120 min; NB30, NB60, and NB120 denote the corresponding no-basal-context forecasts. Circle area represents difference magnitude; blue favors PMB--NN and red favors the comparator. Black outlines and asterisks indicate Holm-adjusted $p<0.05$. B, basal context; NB, no-basal context; MAE, mean absolute error; PMB--NN, physiological-model-based neural network; TCN, temporal convolutional network; GRU, gated recurrent unit; LSTM, long short-term memory; PM, standalone physiological model.
\label{fig:accuracy}
\end{figure*}

The main two-term PMB--NN significantly improved on the standalone PM in all six context--horizon settings and achieved the lowest median MAE at 30 and 60 min in no-basal context (Fig.~\ref{fig:accuracy}; Table~\ref{tab:mae}). Median MAEs were 16.24, 25.00, and 30.20 mg/dL in basal context and 15.38, 24.21, and 31.79 mg/dL in no-basal context at 30, 60, and 120 min, respectively. Error increased with forecast horizon for every model (Fig.~\ref{fig:accuracy}a).

The paired comparisons indicate which differences were consistent across participants (Fig.~\ref{fig:accuracy}b,c). In no-basal context, PMB--NN had significantly lower MAE than TCN, GRU, LSTM, transformer, and PM at both 30 and 60 min. At no-basal 120 min, only PM had significantly higher error than PMB--NN. In basal context, GRU had the lowest median at 30 and 60 min and transformer at 120 min; their paired differences from PMB--NN were not significant. No comparator had a significant MAE advantage over PMB--NN in the six settings.

\begin{table*}[t]
\centering
\caption{Median test mean absolute error (MAE, mg/dL) across participants. Bold marks the lowest median in each row. An asterisk indicates significantly higher comparator error than PMB--NN in a paired two-sided Wilcoxon signed-rank test with Holm correction within each context--horizon setting ($p_{\mathrm{Holm}}<0.05$). PMB--NN, physiological-model-based neural network; TCN, temporal convolutional network; GRU, gated recurrent unit; LSTM, long short-term memory; PM, standalone physiological model.}
\label{tab:mae}
\scriptsize
\resizebox{0.7\textwidth}{!}{%
\begin{tabular}{llrrrrrr}
\toprule
Context & Horizon & PMB--NN & TCN & GRU & LSTM & Transformer & PM \\
\midrule
Basal & 30  & 16.24 & 16.23 & \textbf{15.81} & 17.52 & 16.03 & 18.89$^{*}$ \\
Basal & 60  & 25.00 & 23.66 & \textbf{23.00} & 24.41 & 24.27 & 28.01$^{*}$ \\
Basal & 120 & 30.20 & 30.46 & 30.32 & 31.18 & \textbf{30.12} & 33.53$^{*}$ \\
\midrule
No basal & 30  & \textbf{15.38} & 16.65$^{*}$ & 17.32$^{*}$ & 18.90$^{*}$ & 18.65$^{*}$ & 16.60$^{*}$ \\
No basal & 60  & \textbf{24.21} & 24.46$^{*}$ & 24.77$^{*}$ & 25.23$^{*}$ & 26.10$^{*}$ & 25.72$^{*}$ \\
No basal & 120 & 31.79 & 31.26 & 32.93 & \textbf{30.73} & 31.54 & 33.23$^{*}$ \\
\bottomrule
\end{tabular}%
}
\end{table*}

\subsection{Physiological plausibility}

\begin{figure*}[t]
\centering
\includegraphics[width=\textwidth]{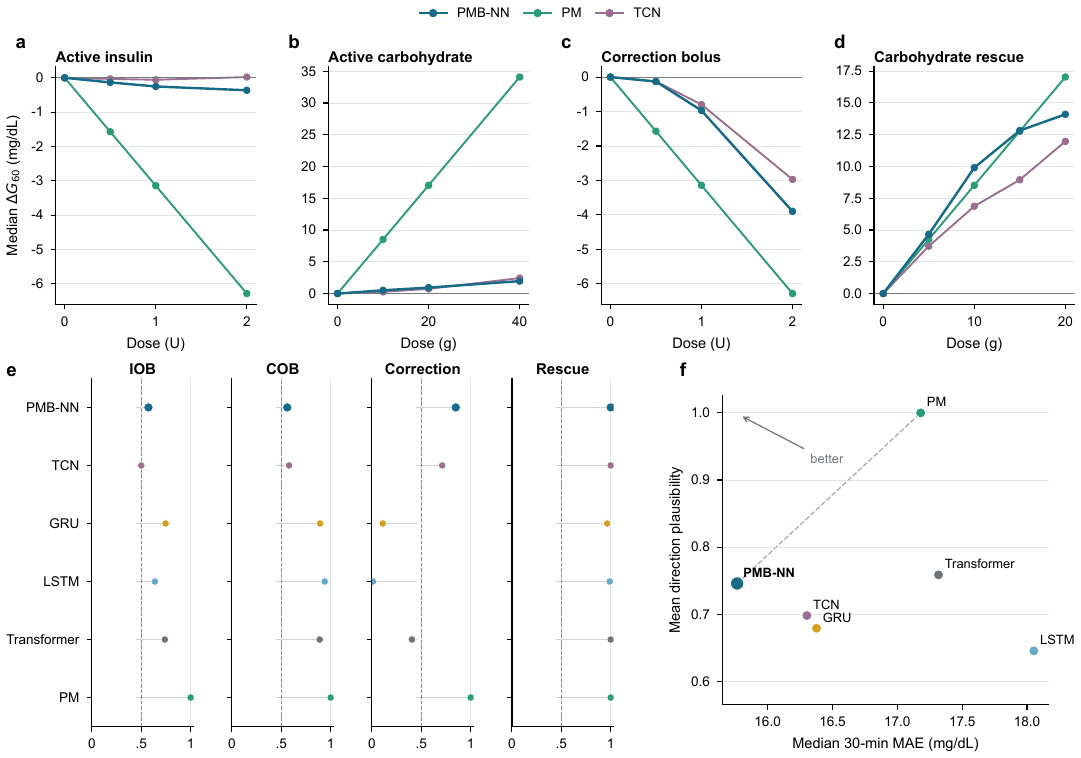}
\caption{Physiological plausibility of the two-term PMB--NN and comparator models under controlled input changes. (a--d) Median changes in 60-min forecasts after increasing IOB, COB, a correction bolus at glucose $\geq180$ mg/dL, or rescue carbohydrate at glucose $\leq80$ mg/dL. Insulin should decrease or leave unchanged the forecast; carbohydrate should increase or leave it unchanged. (e) For each input-change type, the median fraction of eligible test windows with the expected response at the largest dose, summarized across participants, contexts, and horizons. A fraction of 1 indicates that all tested forecasts followed the expected direction. (f) Median 30-min MAE across contexts versus the arithmetic mean of the four model-specific fractions in (e). The upper-left direction represents lower error and more consistent responses. The dashed line joins models for which no displayed comparator has both lower error and higher mean direction plausibility. IOB, insulin on board; COB, carbohydrate on board; MAE, mean absolute error; PMB--NN, physiological-model-based neural network; TCN, temporal convolutional network; GRU, gated recurrent unit; LSTM, long short-term memory; PM, standalone physiological model.}
\label{fig:counterfactual}
\end{figure*}

The two-term PMB--NN followed the expected physiological response more consistently for correction bolus and carbohydrate rescue than for direct changes in active insulin and carbohydrate states (Fig.~\ref{fig:counterfactual}). At 60 min, increasing IOB by 2 U lowered its median forecast by 0.36 mg/dL, and increasing COB by 40 g raised it by 1.96 mg/dL (Fig.~\ref{fig:counterfactual}a,b). These responses were much smaller than those of the standalone PM. In comparison, a 2-U correction bolus lowered the median PMB--NN forecast by 3.90 mg/dL, and a 20-g carbohydrate rescue raised it by 14.10 mg/dL, with larger responses as the event dose increased (Fig.~\ref{fig:counterfactual}c,d).

At the maximum tested doses, median direction-plausible fractions were 0.57 for IOB, 0.56 for COB, 0.85 for correction bolus, and 1.00 for carbohydrate rescue (Fig.~\ref{fig:counterfactual}e). TCN reached 0.71 for correction bolus, while GRU, LSTM, and transformer reached 0.11, 0.01, and 0.41, respectively. All models responded in the expected direction to rescue in nearly every tested record. PM reached 1.00 in all four tests because its equation constrains the response signs. The joint accuracy--plausibility plot shows that PMB--NN combined lower 30-min error with higher mean direction plausibility than TCN, GRU, and LSTM; transformer had slightly higher plausibility but higher error (Fig.~\ref{fig:counterfactual}f).

\subsection{Personalized physiological parameters and their reliability}

\begin{figure*}[h]
\centering
\includegraphics[width=\textwidth]{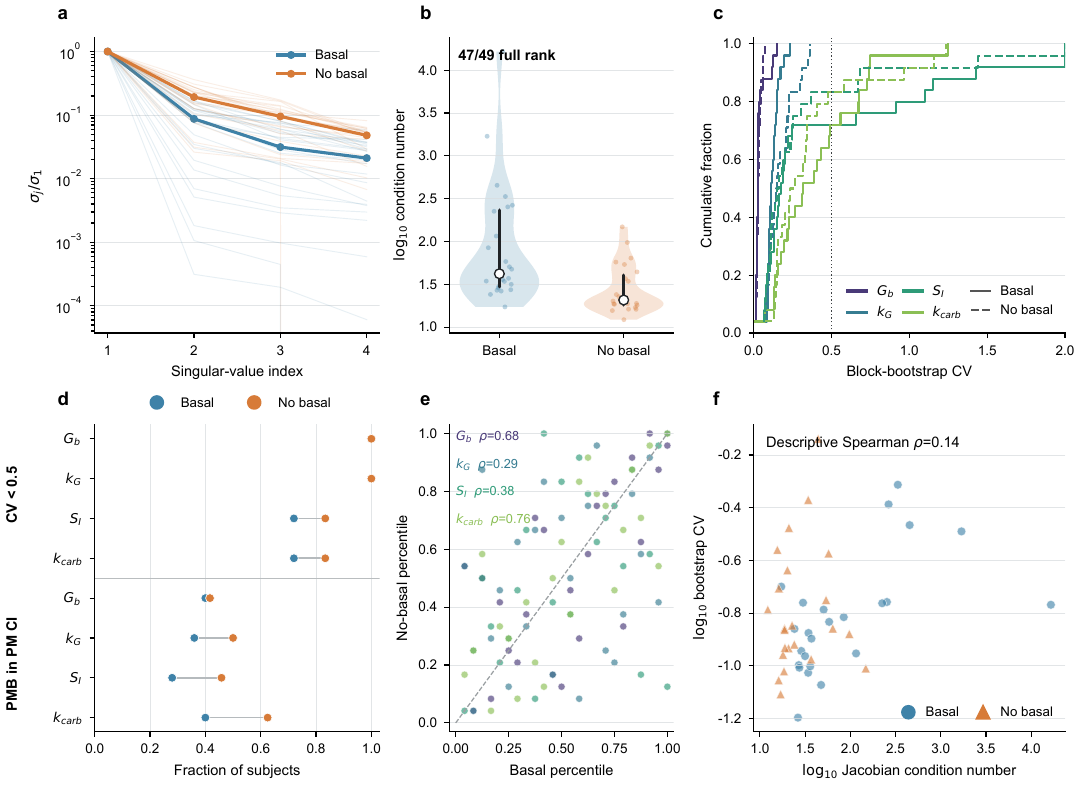}
\caption{Reliability diagnostics for the physiological parameters. Of 49 independent participant--context PM fits, 47 had numerical rank four, i.e., all four parameter effects were locally distinguishable; the other two had rank three. (a) Normalized singular-value spectra of the independent-PM sensitivity Jacobians; thin lines show individual fits and thick lines show context medians. The indices denote directions formed by combinations of scaled parameters, not individual parameters. (b) Condition numbers for the 47 rank-four fits (24 basal and 23 no-basal); black segments and open circles indicate the interquartile range and median. (c) Cumulative distributions of parameter CV from 24-h block-bootstrap PM refits; the dotted line marks CV $=0.5$. (d) Fractions with CV $<0.5$ (upper rows) and fractions of PMB--NN estimates within the independent-PM 95\% bootstrap intervals (lower rows). The latter indicate estimator agreement, not PMB--NN interval coverage. (e) PMB--NN parameter percentiles compared between recording contexts, with Spearman correlations. (f) Relationship between Jacobian conditioning and median parameter CV for the same 47 rank-four fits used in (b). CV, coefficient of variation; CI, confidence interval; PM, physiological model; PMB--NN, physiological-model-based neural network.}
\label{fig:identifiability}
\end{figure*}

The two-term PMB--NN provided participant-specific descriptions of background glucose regulation and input-response strengths, while independent-PM diagnostics showed greater stability for glucose-related parameters than for insulin- and carbohydrate-related coefficients (Table~\ref{tab:parameters}; Fig.~\ref{fig:identifiability}). The median glucose operating point was 167.73 mg/dL with basal records and 156.24 mg/dL without them. Corresponding median return rates were 0.00882 and 0.00752 min$^{-1}$. Table~\ref{tab:parameters} also reports the personalized IOB-based glucose-lowering and COB-based glucose-raising coefficients. These estimates describe effective responses within the reduced equation. The sensitivity and bootstrap results below concern independent physiological fits to the same training data.

The standalone-PM sensitivity Jacobian had full numerical rank in 47 of 49 participant--context fits, indicating distinguishable first-order local effects of the four retained parameters ($G_b$, $k_G$, $S_I$, and $k_{\mathrm{carb}}$) on the forecasts (Fig.~\ref{fig:identifiability}a,b). Among numerically full-rank fits, median condition numbers were 42.3 in basal context and 20.7 in no-basal context. The spread of the normalized singular values indicates unequal forecast sensitivity along different combinations of the range-scaled physical parameters. In the bootstrap analysis, all fits met the CV $<0.5$ stability criterion for $G_b$ and $k_G$, compared with 72.0\% of basal fits and 83.3\% of no-basal fits for each of $S_I$ and $k_{\mathrm{carb}}$ (Fig.~\ref{fig:identifiability}c,d). Agreement between estimators varied by parameter and context: 28.0--62.5\% of PMB--NN estimates fell within the standalone-PM bootstrap intervals (Fig.~\ref{fig:identifiability}d).

Participant rankings across recording contexts were more consistent for $G_b$ ($\rho=0.68$) and $k_{\mathrm{carb}}$ ($\rho=0.76$) than for $k_G$ ($\rho=0.29$) and $S_I$ ($\rho=0.38$; Fig.~\ref{fig:identifiability}e). Across the 47 numerically full-rank fits, Jacobian conditioning had a modest descriptive association with median bootstrap CV (\revmath{\rho=0.28}; Fig.~\ref{fig:identifiability}f). Local parameter separation, resampling stability, and cross-context agreement therefore provided different evidence about parameter interpretation.

\begin{table}[t]
\centering
\caption{Personalized effective physiological parameters estimated by the main two-term PMB--NN. Values are participant median [interquartile range]. $G_b$, glucose operating point; $k_G$, glucose return rate; $S_I$, IOB-based glucose-lowering coefficient; $k_{\mathrm{carb}}$, COB-based glucose-raising coefficient.}
\label{tab:parameters}
\scriptsize
\begin{tabular}{llll}
\toprule
Parameter & Unit & Basal & No basal \\
\midrule
$G_b$ & mg/dL & 167.73 [148.87--174.57] & 156.24 [146.08--166.97] \\
$k_G$ & min$^{-1}$ & 0.00882 [0.00757--0.01000] & 0.00752 [0.00685--0.00884] \\
$S_I$ & mg/dL/min/U & 0.0776 [0.0615--0.0836] & 0.0854 [0.0699--0.0967] \\
$k_{\mathrm{carb}}$ & mg/dL/min/g & 0.02282 [0.01849--0.02984] & 0.01932 [0.01548--0.02478] \\
\bottomrule
\end{tabular}
\end{table}

\subsection{Comparison of two- and three-term PMB--NN formulations}

\begin{figure*}[t]
\centering
\includegraphics[width=\textwidth]{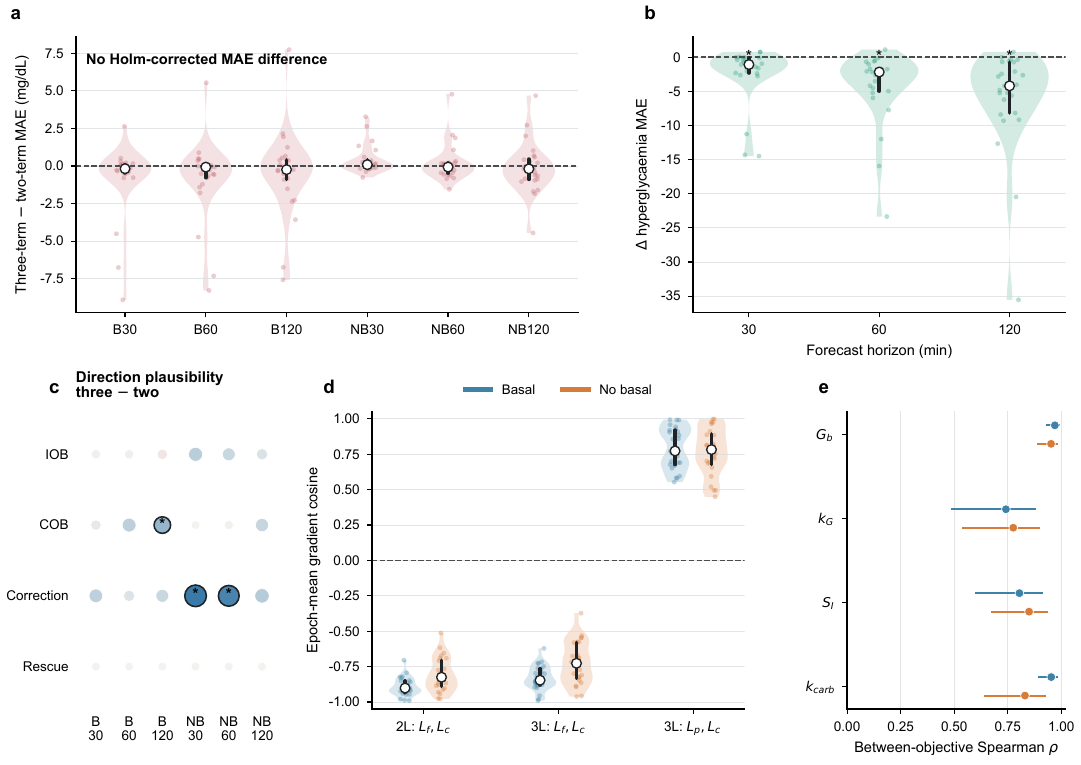}
\caption{Comparison of the main two-term PMB--NN ($L_f+L_c$) and three-term extension ($L_f+0.5L_c+L_p$). (a) Distributions of paired overall MAE changes, defined as three-term minus two-term. Negative values favor the extension. (b) Paired MAE changes for basal-context targets with observed glucose above 180 mg/dL (hyperglycaemia). (c) Median paired changes in direction-plausible fractions at the largest input doses; negative values indicate lower plausibility in the extension. Circle area represents difference magnitude; blue indicates a decrease and red an increase. In (b,c), asterisks indicate Holm-adjusted $p<0.05$. (d) Distributions of epoch-mean gradient cosine similarities; negative values indicate opposing updates within the specified parameter space. (e) Between-formulation Spearman correlations of participant parameter rankings with approximate 95\% intervals. B, basal context; NB, no-basal context; 2L, two-term; 3L, three-term; $L_f$, neural forecast-to-reference loss; $L_c$, neural-to-PM consistency loss; $L_p$, PM forecast-to-reference loss; MAE, mean absolute error; PMB--NN, physiological-model-based neural network.}
\label{fig:formulation_extension}
\end{figure*}

The three-term extension improved accuracy during basal-context hyperglycaemia, with no significant overall MAE difference and lower physiological direction plausibility in three settings (Fig.~\ref{fig:formulation_extension}). None of the six overall MAE comparisons was significant after Holm correction (Fig.~\ref{fig:formulation_extension}a; Supplementary Table~S1). Hyperglycaemia MAE was lower with the extension at all three basal-context horizons ($p_{\mathrm{Holm}}<0.001$; Fig.~\ref{fig:formulation_extension}b). Direction plausibility decreased significantly for basal-context COB at 120 min and no-basal correction bolus at 30 and 60 min (Fig.~\ref{fig:formulation_extension}c).

The loss gradients also differed between formulations (Fig.~\ref{fig:formulation_extension}d). For neural weights, median epoch-mean cosine similarities between $L_f$ and $L_c$ were $-0.90$ and $-0.83$ in the two-term model and $-0.85$ and $-0.73$ in the three-term model for basal and no-basal contexts, respectively. The objectives therefore continued to request opposing neural-weight updates. For physiological parameters in the three-term model, median cosines between $L_p$ and $L_c$ were positive at 0.77 and 0.78. Participant parameter rankings remained similar between formulations, with Spearman correlations of 0.742--0.970 (Fig.~\ref{fig:formulation_extension}e).

\subsection{Short-duration personalized calibration}

Longer calibration generally reduced forecast error, and the full personalization protocol had lower MAE than every 1--15-day setting in basal context (Table~\ref{tab:calibration}). At 1 day, participant-median MAEs were 27.17, 34.37, and 42.41 mg/dL at 30, 60, and 120 min. At 15 days, the corresponding values were 18.29, 25.84, and 33.13 mg/dL, compared with 16.24, 25.00, and 30.20 mg/dL under the full protocol. All 15 paired duration--horizon differences were significant after Holm correction ($p_{\mathrm{Holm}}\leq0.0068$). Error declined more consistently with duration at 30 and 60 min than at 120 min, where the median fluctuated between 6 and 15 days.

The pooled participant--horizon MAE ratio relative to the main model decreased from a median of 1.41 at 1 day to 1.06 at 15 days (Supplementary Fig.~S1b). At 15 days, median absolute relative parameter differences from the full-data estimates were 4\% for $G_b$, 17\% for $k_G$, 11\% for $S_I$, and 12\% for $k_{\mathrm{carb}}$ (Supplementary Fig.~S1d). Participant-rank correlation was 0.93 for $G_b$ and 0.49--0.75 for the other coefficients (Supplementary Fig.~S1e). Supplementary Fig.~S1 shows participant variation and parameter agreement across all five durations.

\begin{table}[t]
\centering
\caption{Basal-context test MAE (mg/dL) after short-duration calibration of the main two-term PMB--NN. Values are participant median [interquartile range] for the same 25 participants and test windows. An asterisk indicates higher error than the full protocol in paired two-sided Wilcoxon tests with Holm correction over the 15 duration--horizon comparisons ($p_{\mathrm{Holm}}<0.05$). The full protocol used the original 70\% training and 10\% validation partitions; its pre-test period had a median duration of 22.2 days (range 20.5--22.5). Short-duration fits retained the final epoch, and the full protocol used validation-selected checkpoints. MAE, mean absolute error; PMB--NN, physiological-model-based neural network.}
\label{tab:calibration}
\scriptsize
\begin{tabular}{llll}
\toprule
Calibration & 30 min & 60 min & 120 min \\
\midrule
1 day & 27.17 [18.81--38.03]$^{*}$ & 34.37 [26.87--51.26]$^{*}$ & 42.41 [31.16--55.90]$^{*}$ \\
3 days & 20.61 [16.98--24.47]$^{*}$ & 29.69 [25.62--34.88]$^{*}$ & 36.59 [32.01--47.91]$^{*}$ \\
6 days & 19.12 [16.57--22.29]$^{*}$ & 28.03 [24.60--31.54]$^{*}$ & 32.90 [30.18--41.34]$^{*}$ \\
10 days & 19.05 [16.85--21.19]$^{*}$ & 27.40 [24.89--32.47]$^{*}$ & 35.96 [29.95--40.16]$^{*}$ \\
15 days & 18.29 [15.67--20.67]$^{*}$ & 25.84 [23.39--30.16]$^{*}$ & 33.13 [26.90--40.39]$^{*}$ \\
\midrule
Full protocol & \textbf{16.24 [13.72--17.77]} & \textbf{25.00 [21.72--27.50]} & \textbf{30.20 [27.06--37.25]} \\
\bottomrule
\end{tabular}
\end{table}

\section{Discussion}
\label{sec:discussion}

This study extends PMB--NN to personalized glucose forecasting by combining a flexible data-driven model with an explicit glucose--insulin--carbohydrate model. Its main value is the joint assessment of numerical accuracy, physiological response behavior, and personalized response coefficients within one framework. The two-term formulation improved on the standalone physiological model and was particularly competitive for 30- and 60-min forecasts without basal records. The three-term extension reduced errors during high-glucose periods with basal records, but did not significantly improve overall MAE and weakened expected responses in three input-test settings. For the configurations evaluated here, this balance favors the two-term formulation as the starting point for personalized forecasting. The parameter diagnostics further distinguish the meaning assigned to each coefficient from the reliability with which it can be estimated.

\subsection{Model performance, personalization, and clinical relevance}
The earlier PMB--NN studies related measured metabolic or pulse-wave inputs to heart-rate and blood-pressure estimation \citep{zhang2026PMBNN1,zhang2026PMBNN2}. The glucose application addresses an additional temporal challenge: insulin and meals have delayed effects, and new intake or treatment can occur after the forecast is issued. The reduced equation represents the expected input effects, while the TCN learns temporal patterns and variation beyond that equation. Their combination improved forecasting over the reduced model alone. The smaller differences among data-driven models in basal context are consistent with recent CGM carrying substantial short-horizon information \citep{martinsson2020blood,kushner2020multi}. The physiological component adds an explicit response structure and person-specific coefficients to this forecasting task.

The potential clinical role is anticipatory warning of rising or falling glucose, complementing alarms based on current CGM. A 30- or 60-min forecast could provide time to check the trend and respond within an existing care plan, while the input-response analysis examines whether the forecast behaves consistently with the intended effect of an insulin or carbohydrate entry. Clinical evaluation should therefore assess warning lead time, missed hypoglycaemia events, false alarms, and rapid glucose changes in addition to average error. Prospective validation is needed to establish the usefulness of these warnings in routine care.

The 30- and 60-min forecasts offer the most relevant starting point for warning-oriented applications. At 120 min, later meals, exercise, and AID actions introduce increasing variation beyond the observed history. This horizon tests the information available to the current model. Longer-horizon forecasting could incorporate repeated updates, uncertainty estimates, or scenario-conditioned predictions based on planned meals and insulin.

The calibration comparison complements the model comparisons by showing how much individual history the data-driven and physiological components had available. More recording days expose the model to a wider range of glucose trajectories and input combinations. The generally smaller errors after longer calibration support this interpretation, while the non-monotonic 120-min results indicate that additional days do not improve every prediction time uniformly. After 15 days of calibration, participants' estimated glucose operating points had a similar ordering to those from the full-data model. Thus relative differences between participants were captured with a shorter recording, even though forecasting remained less accurate than under the full protocol. The other coefficients depended more on the calibration subset. Because the short-duration and full-data protocols also differed in checkpoint selection, a uniform training protocol would isolate the effect of data duration. A short initial calibration followed by periodic updating is a useful direction for prospective personalization studies.

\subsection{Physiological response behavior and parameter interpretation}

The stronger responses to correction bolus and rescue entries suggest that raw events and derived action states have different roles in the learned forecast. An event records a distinct change at the prediction time, whereas IOB and COB summarize earlier delivery through fixed filters. Their weaker responses may reflect redundancy with the preceding glucose trajectory or the way treatment is recorded. In AID data, insulin is often delivered when glucose is already high or rising, creating associations between treatment and subsequent glucose that differ from the isolated action of insulin \citep{pearl2009causality}. Input-response testing reveals how the trained predictor uses these variables. The standalone PM provides a useful reference for expected response signs, and its higher forecast error illustrates the importance of assessing response behavior together with accuracy.

The four fitted coefficients offer a compact description of participant-specific behavior within the chosen glucose model. Their interpretation depends on both the model structure and the variation in the observed inputs. A glucose operating point is supported by many CGM observations, whereas insulin and carbohydrate coefficients depend more directly on the timing and diversity of recorded events. Sparse or co-occurring meals and boluses can make their separate contributions harder to estimate. This provides a plausible explanation for the greater bootstrap variability of the input-related coefficients.

The physiological value of these parameters is to summarize the balance between an individual's background glucose regulation and responses to recorded insulin and carbohydrate over a defined period. The glucose operating point $G_b$ and return rate $k_G$ define the model's background tendency for glucose to return toward $G_b$, with $k_G$ determining the speed of that return. The insulin-response coefficient $S_I$ and carbohydrate-response coefficient $k_{\mathrm{carb}}$ quantify the apparent glucose-lowering strength of insulin on board (IOB) and glucose-raising strength of carbohydrate on board (COB), respectively, under the fixed action filters. This process-based interpretation is distinct from local parameter identifiability and estimation precision. The sensitivity and bootstrap analyses assess these estimation properties, while the coefficient definitions explain what the fitted values mean within the model. With consistent recording and model assumptions, repeated estimates could help compare recording periods and identify changes that warrant closer examination of glucose patterns, delivery, or meal records. Comparison with independently measured response quantities would strengthen their clinical interpretation.

\subsection{Loss formulation and optimization behavior}

The gradient analysis explains how forecast fitting and physiological consistency interact during joint optimization. Negative neural-gradient cosines indicate competing update directions, while positive physiological-gradient cosines in the three-term model indicate alignment between fitting the PM to observations and matching it to the data-driven forecast. The competing objectives are consistent with differences between the patterns represented by the TCN and the processes retained in the reduced equation. Fixed action filters, omitted physiological influences, and treatment-related associations are candidate contributors to this mismatch.

The three-term formulation shifted this balance by fitting the physiological model directly to glucose while reducing the agreement weight. The resulting combination improved errors during high-glucose periods but reduced direction plausibility in three settings. This tradeoff supports the two-term formulation as the main model for the present study and motivates explicit assessment of input responses when choosing loss weights. Independently varying the PM-supervision and consistency weights would help determine how each contributes to subgroup accuracy and physiological behavior.

A related observation was reported in our recent wrist-to-body center-of-mass study \citep{Que2026HybridPhysicsAI}: direct mechanistic supervision generally aligned with the coupling loss, while data-fitting and coupling gradients became weakly aligned or mildly conflicting during training. In our glucose application, strongly negative epoch-mean data-fitting versus consistency cosines occurred in both formulations. The sensor-to-motion mapping and future-glucose forecasting task differ in their mechanistic approximations, available inputs, and unobserved influences. These differences provide hypotheses for the distinct learning behavior and motivate coupling strategies that are evaluated within each application.

\subsection{Limitations and future directions}

Evaluation covered 25 mostly older participants from one pump--CGM ecosystem, with one training seed for the data-driven models. External cohorts and repeated optimization would assess reproducibility across populations, devices, and initializations. The basal and no-basal contexts represent different recording periods with different available inputs; a matched-input analysis within the same period would help separate these influences. The reduced PM summarizes glucose, insulin, and carbohydrate effects, while activity, illness, meal composition, and future treatment remain additional sources of variation. Parameter uncertainty was evaluated for the independently fitted PM; extending this analysis to PMB--NN requires resampling and refitting both jointly trained modules. Likewise, input perturbations assess the trained forecast's response, while intervention studies are needed to determine actual treatment effects. These are priorities for broader validation of personalized forecasting.

\section{Conclusion}

PMB--NN combines personalized glucose forecasting with explicit insulin and carbohydrate effects and model-based response parameters. The two-term formulation improved on the standalone physiological model and had lower errors than neural comparators at 30 and 60 min without basal records. For the evaluated configurations, it offered the preferred balance of overall accuracy and physiological response consistency: the three-term extension improved high-glucose forecasting with basal records but did not significantly improve overall MAE and reduced plausibility in three input-test settings. With external and prospective validation, the two-term PMB-NN has potential to support interpretable short-horizon glucose warnings.

\section*{Data availability statement}

The AZT1D dataset analyzed in this study is publicly available from Mendeley Data at \href{https://doi.org/10.17632/gk9m674wcx.1}{https://doi.org/10.17632/gk9m674wcx.1}.

\section*{Ethics statement}

The present study was a secondary analysis of de-identified, publicly available data and involved no new participant recruitment or intervention.

\section*{Author contributions}

YZ: Conceptualization, Methodology, Software, Formal analysis, Investigation, Validation, Visualization, Writing--original draft, and Writing--review and editing. SQ: Methodology and Writing--review and editing. YW: Conceptualization, Methodology, Project administration and Writing--review and editing. All authors approved the submitted version and accept responsibility for the work.

\section*{Funding}

The authors received no specific funding for this work.

\section*{Conflict of interest}

The authors declare that the research was conducted in the absence of any commercial or financial relationships that could be construed as a potential conflict of interest.

\section*{Generative AI statement}

The authors declare that generative AI was used to assist with language editing, LaTeX formatting, and the preparation of analysis and visualization code. The authors take full responsibility for the accuracy and integrity of the analyses and final manuscript.

\bibliographystyle{Frontiers-Harvard}
\bibliography{main}
\end{document}